\documentclass[twocolumn, table]{aastex631}
\usepackage{color}
\usepackage{graphicx}
\usepackage{amsmath}
\usepackage{subfigure}

\newcommand{\rmsub}[1]{\ensuremath{_{\mathrm{#1}}}}

\newcommand{\yr}{\textrm{yr}} 
\newcommand{\ps}{\ensuremath{\mathrm{s^{-1}}}} 
\newcommand{\km}{\textrm{km}} 
\newcommand{\pc}{\textrm{pc}} 

\newcommand{\Msun}{\ensuremath{\mathrm{M_{\Sun}}}} 
\DeclareUnicodeCharacter{02BC}{'}

\shorttitle{Thirty-five years of timing of M53A}
\shortauthors{Lian, et al}

\begin{document}

\title{Thirty-five years of timing of M53A with Arecibo and FAST}

\author{Yujie Lian}
\affiliation{Institute for Frontiers in Astronomy and Astrophysics, Beijing Normal University, Beijing 102206, Peopleʼs Republic of China}
\affiliation{School of Physics and Astronomy, Beijing Normal University, Beijing 100875, Peopleʼs Republic of China}
\affiliation{Max-Planck-Institut f\"ur Radioastronomie, Auf dem H\"ugel 69, D-53121 Bonn, Germany}

\author{P. C. C. Freire$^{\ast}$}
\affiliation{Max-Planck-Institut f\"ur Radioastronomie, Auf dem H\"ugel 69, D-53121 Bonn, Germany}

\author{Shuo Cao$^\dag$} 
\affiliation{Institute for Frontiers in Astronomy and Astrophysics, Beijing Normal University, Beijing 102206, Peopleʼs Republic of China}
\affiliation{School of Physics and Astronomy, Beijing Normal University, Beijing 100875, Peopleʼs Republic of China}

\author{Mario Cadelano}
\affiliation{Dipartimento di Fisica e Astronomia ``Augusto Righi,'' Alma Mater Studiorum Universit\`a di Bologna, via Piero Gobetti 93/2, I-40129 Bologna, Italy}
\affiliation{INAF-Osservatorio di Astrofisica e Scienze dello Spazio di Bologna, Via Piero Gobetti 93/3, I-40129 Bologna, Italy}

\author{Cristina Pallanca}
\affiliation{Dipartimento di Fisica e Astronomia ``Augusto Righi,'' Alma Mater Studiorum Universit\`a di Bologna, via Piero Gobetti 93/2, I-40129 Bologna, Italy}
\affiliation{INAF-Osservatorio di Astrofisica e Scienze dello Spazio di Bologna, Via Piero Gobetti 93/3, I-40129 Bologna, Italy}

\author{Zhichen Pan}
\affiliation{National Astronomical Observatories, Chinese Academy of Sciences, 20A Datun Road, Chaoyang District, Beijing, 100101, Peopleʼs Republic of China}
\affiliation{Guizhou Radio Astronomical Observatory, Guizhou University, Guiyang 550025, Peopleʼs Republic of China}
\affiliation{Key Laboratory of Radio Astronomy and Technology, National Astronomical Observatories, Chinese Academy of Sciences, Beijing 100101, Peopleʼs Republic of China}
\affiliation{College of Astronomy and Space Sciences, University of Chinese Academy of Sciences, Beijing 100049, Peopleʼs Republic of China}

\author{Haiyan Zhang$^{\ddag}$}
\affiliation{National Astronomical Observatories, Chinese Academy of Sciences, 20A Datun Road, Chaoyang District, Beijing, 100101, Peopleʼs Republic of China}
\affiliation{Key Laboratory of Radio Astronomy and Technology, National Astronomical Observatories, Chinese Academy of Sciences, Beijing 100101, Peopleʼs Republic of China}
\affiliation{College of Astronomy and Space Sciences, University of Chinese Academy of Sciences, Beijing 100049, Peopleʼs Republic of China}

\author{Baoda Li}
\affiliation{College of Physics, Guizhou University, Guiyang 550025, Peopleʼs Republic of China}

\author{Lei Qian$^{\S}$}
\affiliation{National Astronomical Observatories, Chinese Academy of Sciences, 20A Datun Road, Chaoyang District, Beijing, 100101, Peopleʼs Republic of China}
\affiliation{Guizhou Radio Astronomical Observatory, Guizhou University, Guiyang 550025, Peopleʼs Republic of China}
\affiliation{Key Laboratory of Radio Astronomy and Technology, National Astronomical Observatories, Chinese Academy of Sciences, Beijing 100101, Peopleʼs Republic of China}
\affiliation{College of Astronomy and Space Sciences, University of Chinese Academy of Sciences, Beijing 100049, Peopleʼs Republic of China}

\footnote{$\ast$ pfreire@mpifr-bonn.mpg.de}
\footnote{$\dag$ caoshuo@bnu.edu.cn}
\footnote{$\ddag$ hyzhang@bao.ac.cn}
\footnote{$\S$ lqian@bao.ac.cn}

\begin{abstract}

PSR B1310+18A is a 33-ms binary pulsar in a 256-day, low eccentricity orbit with a low-mass companion located in NGC 5024 (M53). In this Letter, we present the first phase-coherent timing solution for this pulsar (designated as M53A) derived from a 35-year timing baseline; this combines the archival Arecibo Observatory data with the recent observations from the Five-hundred-meter Aperture Spherical radio Telescope (FAST). We find that the spin period derivative of the pulsar is between 6.1 and $7.5 \times 10^{-19} \rm \, s\, s^{-1}$, which implies a characteristic age between 0.70 and 0.85 Gyr. The timing solution also includes a precise position and proper motion for the pulsar, enabling the identification of the companion of M53A in Hubble Space Telescope data as a Helium white dwarf (He WD) with a mass of $M_{\rm WD}=0.39^{+0.05}_{-0.07} \, \Msun$ and a cooling age of $0.14^{+0.04}_{-0.03}\, \rm Gyr$, confirming that the system formed recently in the history of the GC. The system resembles, in its spin and orbital characteristics, similarly wide pulsar - He WD systems in the Galactic disk. We conclude by discussing the origin of slow pulsars in globular clusters, showing that none of the slow pulsars in low-density globular clusters are as young as the systems observed in the densest known globular clusters.

\end{abstract}


\section{Introduction}
\label{sec:introduction}

Globular clusters (GCs) are characterized by exceptionally dense stellar environments ($10^3 - 10^6\, \rm pc^{-3}$), where interactions and collisions between stars are frequent over their lifetimes. 
As a result, these clusters host an unusually high number of low-mass X-ray binaries (LMXBs) and their evolutionary descendants, millisecond pulsars (MSPs), relative to their stellar mass \citep{Clark1975, Katz1975}. The reason for this is that, unlike in the Galactic disk, LMXBs in GCs typically form through tidal capture of another star by neutron stars (NSs, e.g., \citealt{1975MNRAS.172P..15F}) and exchange interactions, where a NS replaces one member of a primordial binary system during a close stellar encounter \citep{Sigurdsson1993}. The evolutionary pathways of LMXBs in clusters lead to diverse outcomes, particularly in the globular clusters with more frequent stellar interactions \citep{Sigurdsson1995,Verbunt2014}.
A consequence of the evolution of the GC from LMXBs is that binary pulsars account for $\sim$ 56\% among all 344 pulsars reported in 45 GCs\footnote{\url{https://www3.mpifr-bonn.mpg.de/staff/pfreire/GCpsr.html}}, a stark contrast to the $\sim$ 12\% binary fraction observed in the Galaxy as a whole \citep{Manchester2005}\footnote{\url{https://www.atnf.csiro.au/research/pulsar/psrcat/}}.

The population of binary pulsars in globular clusters has, compared to that of the Galaxy, fewer wide systems: the number of systems with orbital period $P_{\rm b} < 50$\,d comprising $\sim$ 96\% of the total binary population. Part of the reason is the high stellar densities in many GCs, which destabilize wide systems on relatively short timescales.
Only 6 long-period binaries ($P_{\rm b} > 50$\,d) in GCs have been reported, and these generally have mild eccentricities $\sim 10^{-2}$, except for PSR J1748$-$2446ao, a possible DNS in an eccentric orbit ($e=0.32$; \citealt{Padmanabh2024}) in the globular cluster Terzan 5, that very likely formed in an exchange encounter that happened after the pulsar as fully recycled, which is something that is more likely to happen in dense GCs like Terzan 5. However, simulations suggest that $\sim 30\%$ binaries with $P_{\rm b} > 50$\,d may survive after 13 Gyr evolution in GCs that have low density ($\sim 10^{3}\rm \, \Msun \; pc^{-3}$), small velocity dispersion ($\sigma_{v} = 5 \, \rm km\,s^{-1}$), and initial binary fraction of 50\% \citep{Sollima2008}.

NGC 5024 (M53, $\alpha = 13^{\mbox{\scriptsize h}} 12^{\mbox{\scriptsize m}} 55.3^{\mbox{\scriptsize s}}$, $\delta = +18\degr 10\arcmin 05\arcsec$) 
is an old (age $\approx$ 13\,Gyr, \citealt{ForbesBridges2010}) located in the intermediate Galactic halo at a Heliocentric distance of $d = 18.5$\,kpc \citep{Baumgardt2021}, making it the most distant GC with known pulsars. 
With a central density of $\rho_c \sim 1.2\times 10^3 \,\rm L_{\odot}\,pc^{-3}$, M53 is the second least dense GC hosting pulsars \citep{Harris1996,Harris2010}, behind only M71 (\citealt{Cadelano2015}, Lian et al, in prep.).
Low-density clusters like M53 are particularly favorable for detecting long-orbit pulsars, as reduced stellar interactions minimize the possibility of binary disruption.
PSR B1310+18A is a 33.16\,ms pulsar, discovered in 1989 with the Arecibo 305-m radio telescope \citep{Anderson1989} towards M53. We will refer to it as M53A. It was one of the very earliest known pulsars in a GC. Subsequent follow-up revealed that it is a member of a binary system, with an orbital period $P_\mathrm{b}\, \sim \, 256 \, \rm d$ and low orbital eccentricity
\citep[$e < 0.01$,][]{Kulkarni1991}. The projected semi-major axis of $x = 82.4(7)$\,lt-s 
indicates, assuming a pulsar mass of 1.4\,$\Msun$, a minimum companion mass of 0.305\,$\Msun$. This system is remarkable in two ways: it is still in the most distant GC with known pulsars, and until the recent discoveries of PSR 1953+1846B (M71B) and PSR 1953+1846C (M71C), its orbital period was the largest known among GC pulsars. However, despite this significant follow-up effort, no phase-connected timing solution was published in the following 34 years.

As one of the essential targets for searching pulsars in the Five-hundred-meter Aperture Spherical radio Telescope (FAST; \citealt{Nan2011,Jiang2020}) GC pulsar survey, \textbf{G}lobular \textbf{C}luster with \textbf{F}AST: \textbf{A} \textbf{N}eutron-star \textbf{S}urvey (GC FANS\footnote{\url{https://fast.bao.ac.cn/cms/article/65/}}), we have found four new MSPs (PSRs J1312+1810B, C, D and E, henceforth M53B to E) by FAST \citep{Pan2021,Lian2023}. Among new discoveries, M53B, D, and E are all in a binary system and have mild eccentricities ranging from $10^{-2}$ to $10^{-5}$, as well as low-mass companions ranging from 0.18 to 0.27 $\Msun$, small magnetic fields ($\sim 10^9 \, \rm G$) and large characteristic ages ($> 2 \, \rm Gyr$). The pulsar population in M53 thus resembles the MSP population observed in the Galactic disk \citep{Manchester2005}. In this Letter, we present the phase-coherent timing solution for M53A from the early 1990s to the current time using archival Arecibo data and data from ongoing observation with FAST. In Section~\ref{sec:observations}, we describe the observations and data reduction. In Section~\ref{sec:timing results}, we present the results of the timing analysis of M53A. The optical analysis of M53A in archival Hubble Space Telescope (HST) observations is presented in Section~\ref{sec:optical counterpart}.  We discuss and summarize our findings in Section~\ref{sec:conclusions}.

\section{Observations and Data reduction} 
\label{sec:observations}

\subsection{FAST and Arecibo observations}

For FAST observations, M53 was initially tracked on November 30$^{th}$ 2019, as the pilot survey for GC FANS \citep{Pan2021}.
We carried out 36 FAST observations from 2019.11 to 2024.05, using the central beam of the FAST 19-beam L-band receiver, which has a beam size of $\rm \sim 3 ^\prime$ and covers a frequency range of 1.0-1.5\,GHz.
All the FAST observations were 8-bit sampled for two polarizations and channelized into 4096 channels (0.122 MHz channel width), the resulting power spectra were summed every 49.152 $\mu \rm s$.
The search of M53A was done on all of the data with the PulsaR Exploration and Search TOolkit \citep[{\sc presto}, ][]{Ransom2001,Ransom2002,Ransom2003}.
We then derived times of arrival (ToAs) by cross-correlating the pulse profiles against a high signal-to-noise ratio (S/N) template, which was obtained from fitting a set of Gaussian curves to the best detections.
For the subsequent analysis of the ToAs, we used {\sc tempo} pulsar timing package \citep{Nice2015} \footnote{\url{http://tempo.sourceforge.net}}. 

For Arecibo observations, M53 was first observed by the 430\,MHz Carriage House line-feed receiver during 1989.03 and 1993.09, using the Arecibo digital correlator as a back-end \citep{Kulkarni1991}. The uncertainty estimates for the ToAs derived from the above observations were not estimated\footnote{The ToAs of M53A from these observations, which were until now unpublished, were provided by the observers: Alex Wolszczan, Stuart B. Anderson, Bryan Jacoby, and Shrinivas Kulkarni.}, for this reason we added a time constant in quadrature, in such a way that the reduced $\chi^2$ for this data set is 1.0. Arecibo resumed the observations on M53 during 2003.07 and 2008.02 using the Gregorian L-band receiver at 1175\,MHz and 1475\,MHz.
The observing details are very similar to those discussed by \cite{Freire2008}.
In these observations, only M53A is detectable. The much weaker signals of M53B to E, compared to M53A, make them difficult to detect in the Arecibo data; for this reason we could not extend their timing solutions to the early 1990s using the Arecibo data.

\subsection{Hubble Space Telescope}

In this work we used ultraviolet (UV) and optical data obtained with the UVIS channel of the Wide Field Camera 3 on HST, using three different filters: F275W, F336W, and F438W, obtained as part of the HST proposal: GO-13297 (PI: Piotto).
The observations of M53 were conducted in two epochs: 2013.12 and 2014.03.
Each epoch included three images per filter, with exposure times of 1733\,s for F275W, 433\,s for F336W, and 170\,s for F438W.
Source detection and PSF photometry was performed with {\sc DAOPHOT IV} on calibrated images, following the ``UV-route'' approach. 
Details of the data reduction process can be found in \citet{Chen2021} and \citet{Cadelano2020}.
The photometry was calibrated to the VEGAMAG system using appropriate zero-points and aperture corrections \citep{Piotto2015}. 
Following \citet{Bellini2011}, source positions were aligned to the International Celestial Reference System (ICRS) by cross-matching with Gaia DR3 \citep{GaiaCollaboration2023}, achieving a 1$\sigma$ astrometric accuracy with a root-mean-square residual of about 17\,mas.

\begin{table}
\begin{center}
\caption{Observed and derived parameters of M53A. The distance of M53 to the Sun is assumed to be 18.5\,kpc to calculate the offset of M53A to the center (labeled by $s$).}
\label{table:timingM53A}
\resizebox{\columnwidth}{!}{
\begin{tabular}{l c}
\hline\hline
Pulsar  &   1312+1810A                                                             \\
Reference Epoch (MJD)                        &   60208.000000                                           \\
Start of Timing Data (MJD)                   &   47666.079                                            \\
End of Timing Data (MJD)                    &   60457.678                                             \\
Number of TOAs           &   5244                                                                    \\
EFAC of Arecibo data &      1.16                                 \\
EFAC of FAST data &         1.30                              \\
Residuals RMS ($\mu$s)    &   75.60                                        \\
Reduced $\chi^2$   &     1.001                                  \\
Solar System Ephemeris       &   DE440                                                                  \\
Binary Model          &   DD                                                                     \\
\hline
\multicolumn{2}{c}{Measured Quantities}  \\
\hline
Right Ascension, $\alpha$ (J2000)             & 13:12:53.67911(8)     \\
Declination, $\delta$ (J2000)                 & 18:10:27.580(2)                                     \\
Proper motion in $\alpha$, $\mu_{\alpha}$ (mas\,yr$^{-1}$)   & $-$0.35(15)     \\
Proper motion in $\delta$, $\mu_{\delta}$ (mas\,yr$^{-1}$)   & $-$0.61(22)     \\
Spin Frequency, $f_{\rm 0}$ (s$^{-1}$)        &   30.153934832592(3)                                   \\
$1^{\rm st}$ Spin Frequency derivative, $f_{\rm 1}$ (Hz s$^{-2}$)   &   $-$6.1199(5)$\times 10^{-16}$ \\
$2^{\rm st}$ Spin Frequency derivative, $f_{\rm 2}$ (Hz s$^{-3}$)   &   $-$1.08(10)$\times 10^{-27}$ \\
Dispersion Measure, DM (pc cm$^{-3}$)        &    24.931(2)                                         \\
Projected Semi-major Axis, $x_{\rm p}$ (lt-s)   &   84.178615(2)                         \\
Orbital Eccentricity, $e$    &   0.00055732(5)                      \\
Longitude of Periastron, $\omega$ (deg) &   139.803(5)                                   \\
Epoch of passage at Periastron, $T_0$ (MJD)  &   60208.741(4)                   \\
Orbital Period, $P_{\rm b}$ (days)    &   255.85737271(9)       \\
Rate of change of $x$, $\dot{x}$ ($10^{-12}$\,lt-s\,s$^{-1}$)  &   $-$0.022(7)   \\
\hline
\multicolumn{2}{c}{Derived Quantities}  \\
\hline
Spin Period, $P$ (s)     &   0.033163167777332(3)                                   \\
1st Spin Period derivative, $\dot{P}$ (s s$^{-1}$)  &    6.7307(6)$\times 10^{-19}$         \\
Mass function, $f(M_{\rm p}, M_{\rm c})$ ($M_{\rm \odot}$) &  0.0097830570(5) \\
Minimum companion mass, $M_{\rm c, min}$ ($\Msun$) &  0.3053 \\
Angular offset from center in $\alpha$, $\theta_{\alpha}$ (arcmin)  &   -0.3731  \\
Angular offset from center in $\delta$, $\theta_{\delta}$ (arcmin)  &   0.3697 \\
Total angular offset from center, $\theta_{\perp}$ (arcmin)         &   0.5253 \\ 
Total angular offset from center, $\theta_{\perp}$ (core radii)     &   1.5007   \\ 
Projected distance from center, $r_{\perp}$ (pc)                    &   2.8266$^s$ \\
\hline
\end{tabular}}
\end{center}
\end{table}

\section{Timing results}
\label{sec:timing results}

\begin{figure*}
	\begin{center}
		\includegraphics[width=0.8\textwidth,height=0.4\textwidth]{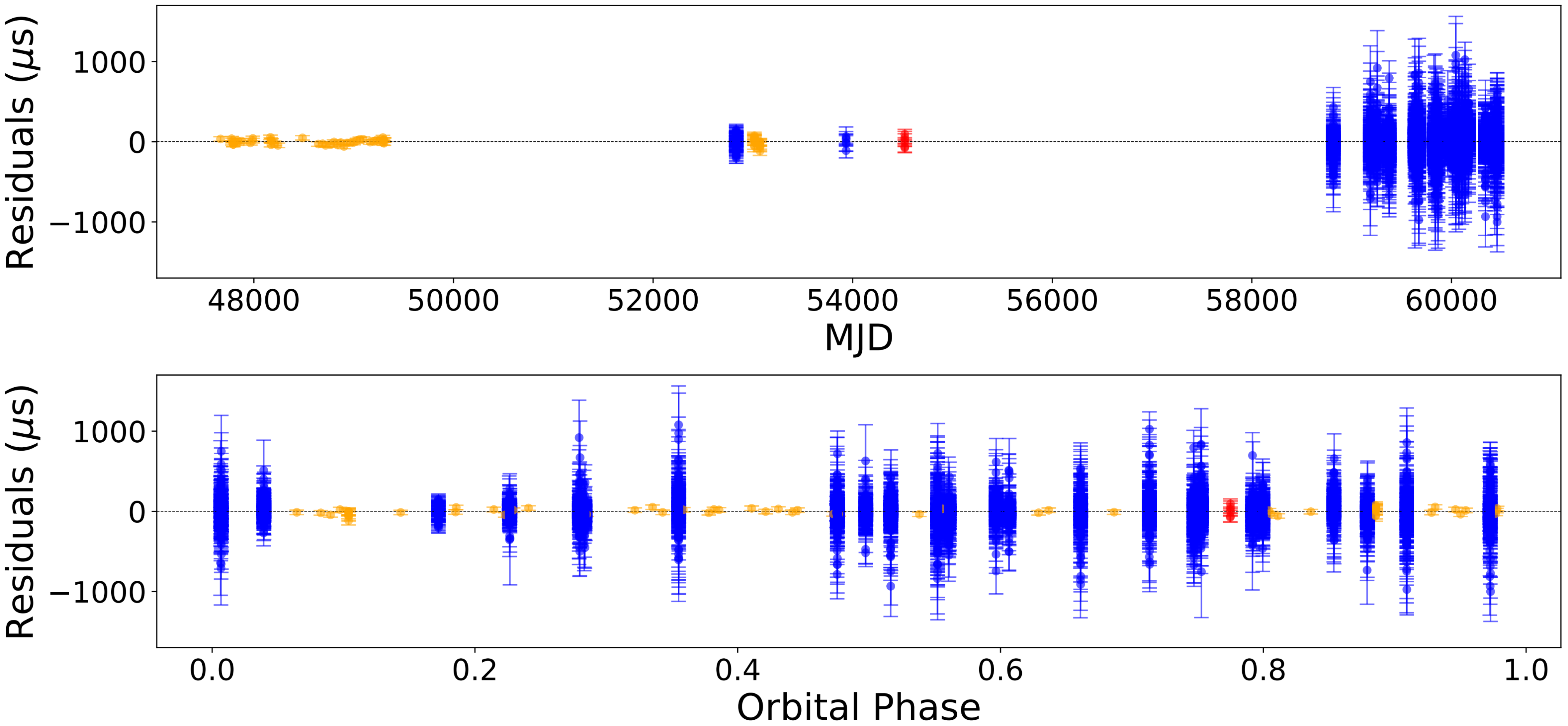}
	\end{center}
	\caption{Timing residuals from the best-fit timing models presented in Table~\ref{table:timingM53A} as a function of the observation date for M53A.}
        \label{residual}
\end{figure*}

Our timing solution of M53A, which spans over 35 years, is shown in Table~\ref{table:timingM53A}, while Fig.~\ref{residual} displays the post-fit timing residuals (ToA minus prediction of the ephemeris for that rotation) over time and the orbital phase. The absence of trends in the residuals shows that no noticeable unmodeled effects are apparent in the data. In what follows, we discuss some of these timing parameters.

\subsection{Position and Proper Motion}
\label{sec:astrometry}

The timing solution includes an accurate position of M53A: $\alpha = 13^{\mbox{\scriptsize h}} 12^{\mbox{\scriptsize m}} 53.67911^{\mbox{\scriptsize s}}$(8), $\delta = +18\degr 10\arcmin 27.580\arcsec$(2), at the reference epoch. At an angular distance  $\theta_{\perp} = 0.52$ arcminutes from the center of the cluster, the pulsar lies outside the cluster core, which has a radius of $r_{\rm c}=0.35\arcmin$ \citep{Harris2010}, thus $\theta_{\perp} \sim 1.5r_{\rm c}$.

The pulsar's proper motion is $\mu_{\alpha}\, =\, -0.35 \pm 0.15 \, \rm mas\,yr^{-1}$ and $\mu_{\delta}\, =\, -0.61 \pm 0.22 \, \rm mas\,yr^{-1}$. The values for M53 are $\mu_{\alpha, \rm M53}\, =\, -0.133 \pm 0.024\, \rm mas\,yr^{-1}$ and $\mu_{\delta, \rm M53}\, =\, -1.331 \pm 0.024\, \rm mas\,yr^{-1}$ (from Gaia DR3; \citealt{Vasiliev2021}). The difference is therefore $\mu_{\alpha} - \mu_{\alpha, \rm M53} = -0.22 \pm 0.15 \, \rm mas\,yr^{-1}$, which is 1.4-$\sigma$ significant, and $\mu_{\delta} - \mu_{\delta, \rm M53} = 0.72 \pm 0.22 \, \rm mas\,yr^{-1}$, which is 3.3-$\sigma$ significant. This difference is very large: at a distance of 18.5 kpc, the escape velocity of M53 (25.9\,km s$^{-1}$\footnote{\url{https://people.smp.uq.edu.au/HolgerBaumgardt/globular/parameter.html}}) translates into a difference of proper motions of at most 0.3~mas~yr$^{-1}$. This suggests that our measurement of the proper motion might be affected by correlations with other parameters, or perhaps some unknown systematics in our timing, like DM variations. For this reason, we will assume from now on that the proper motion of the pulsar is the same as for M53. The precise position and proper motion allowed the optical identification of the companion, which is described in Section~\ref{sec:optical counterpart}.

\subsection{Spin Period and orbital period derivatives}
\label{sec:period_derivatives}

\begin{figure}
	\includegraphics[width=\columnwidth]{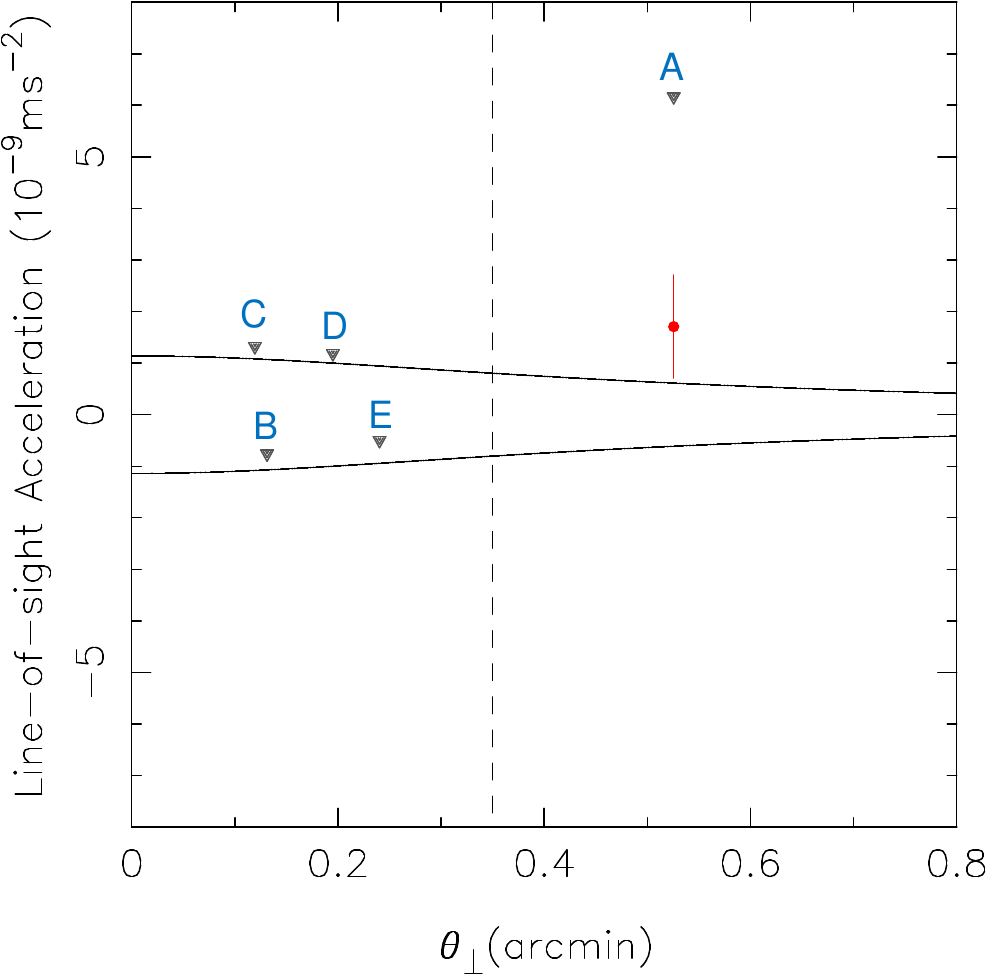}
	\caption{Acceleration model for M53. The black solid lines represent the upper and lower limits for the LOS accelerations ($a_{\ell \rm GC}$) caused by the cluster as a function of the total angular offset from the centre of the cluster ($\theta_{\perp}$). The triangles pointing down represent independent upper limits for the pulsar accelerations along the line of sight, $a_{\ell, \rm P, max}$ for M53A (this Letter) and the remaining pulsars in M53 \citep{Lian2023}. The vertical dashed line is the core radius. The cluster model can account for the negative $\dot{P}$ values of M53B and M53E, but can not fully explain the $a_{\ell, \rm P, max}$ of M53A, C and D, this positive difference is likely due to the intrinsic spin-down of these pulsars. The red error bar shows the measurement of the LOS acceleration of M53A derived from its orbital period derivative, $\dot{P}_{\rm b,obs}$, confirming that it has a LOS acceleration that is consistent with the cluster predictions, and confirming the large intrinsic spin-down of M53A.}
        \label{accel}
\end{figure}

We now discuss the measurements of the first and second spin period derivatives of M53A. In Table~\ref{table:timingM53A}, we see that M53A has a positive $\dot{P}$. Radio pulsars are generally spin-powered, so their intrinsic spin period derivative $\dot{P}_{\rm int}$ should be positive. However, there are additional contributions to the observed spin period derivative ($\dot{P}_{\rm obs}$)
\begin{equation}
\left( \frac{\dot{P}}{P} \right)_{\rm obs} = \left( \frac{\dot{P}}{P} \right)_{\rm int} + \frac{\mu^2d}{c} + \frac{a_{\ell,\, \rm GC}}{c} + \frac{a_{\rm Gal}}{c}.
\label{eq:spinderivative}
\end{equation}
The second term on the right represents the so-called Shklovskii effect (see \citealt{Shklovskii1970}), which depends on the total proper motion of the pulsar (assumed to be the same as that of M53, see Section~\ref{sec:astrometry}) and the distance to M53 ($d = 18.5$ kpc), yielding an acceleration of $0.024 \times 10^{-9}\, \rm m\,s^{-2}$.
The term $a_{\rm Gal}$ describes the line-of-sight (LOS) difference in the Galactic acceleration between M53's center of mass and that of the solar system barycenter, calculated here as $a_{\rm Gal}=-0.105 \times 10^{-9}\, \rm m\,s^{-2}$ using the Galactic potential model of \citet{McMillan2017}. This provides an improved description of the Galactic potential for high Galactic latitudes (for M53, $b = 79^\circ$) relative to the analytical model used by \cite{Lian2023}. The term $a_{\ell,\, \rm GC}$ represents the LOS acceleration due to the cluster's gravitational field. In this Letter, we use the same model of the gravitational field of M53 described by \cite{Lian2023}, which is based on the analytical model discussed by \cite{Freire2005}.
In Fig.~\ref{accel}, the solid black curves denote the maximum and minimum values of the acceleration caused by the gravitational field of the cluster at each angular distance from the center, $\theta_{\perp}$.

For each pulsar in M53, we can derive an independent upper limit on the cluster acceleration from $\dot{P}_{\rm obs}$ (indicated as triangles pointing down, see Fig.~\ref{accel})
\begin{equation}
a_{\ell, \rm P, max} \, = \, c \frac{\dot{P}_{\rm obs}}{P}\, - \, \mu^2 d \, - \, a_{\rm Gal} = 6.17 \, \times \, 10^{-9} \, \rm m \, s^{-2},
\end{equation}
which assumes $\dot{P}_{\rm int} = 0$. The small line-of-sight accelerations predicted by our mass model of M53 ($| a_{\ell,\, \rm GC}| \leq 0.61 \times 10^{-9}\,\rm m\,s^{-2}$) can not account for the positive $ a_{\ell, \rm P, max}$ of M53A, indicating that its intrinsic acceleration $\dot{P}_{\rm int}$ dominates the positive $\dot{P}_{\rm obs}$. Using Eq.~(\ref{eq:spinderivative}) and subtracting the maximum and minimum accelerations caused by the GC potential, we obtain the minimum and maximum limits for $\dot{P}_{\rm int}$: 6.15 and 7.50$\times 10^{-19}\,\rm s\,s^{-1}$ respectively. This corresponds to a characteristic age of 0.70 - 0.85\,Gyr and a magnetic field strength of 4.55 - 5.03$\times 10^9\,\rm G$. From Fig.~\ref{accel}, we can conclude that the $\dot{P}_{\rm int}/P$ of M53A, is significantly larger than for the other pulsars in M53 \citep{Lian2023}.

For each binary in GC, the LOS acceleration terms can also be independently obtained from the observed orbital period derivative \citep{Damour1991}
\begin{equation}
\left( \frac{\dot{P}_{\rm b}}{P_{\rm b}} \right)_{\rm obs} = \left( \frac{\dot{P}_{\rm b}}{P_{\rm b}} \right)_{\rm int} + \frac{\mu^2d}{c} + \frac{a_{\ell,\, \rm GC}}{c} + \frac{a_{\rm Gal}}{c},
\label{eq:orbitalderivative}
\end{equation}
where the last three terms are listed in Eq.~(\ref{eq:spinderivative}).
$\dot{P}_{\rm b, int}$ is the intrinsic orbital period derivative, which is the energy loss due to the emission of gravitational waves and is normally negligible for wide pulsar-WD systems; in this case its expected value is of the order of $10^{-18} \, \rm s\,s^{-1}$, 8 orders of magnitude smaller than the measurement uncertainty.

Fitting the orbital period derivative of M53A yields $\dot{P}_{\rm b,obs} = 1.19 \pm 0.74 \times 10^{-10} \, \rm s\,s^{-1}$, which is not significant but provides some constraints on $a_{\ell,\, \rm GC}$ (red bar in Fig.~\ref{accel}); these are 1-$\sigma$ consistent with the predictions of the cluster model. Thus, we can rule out significant contributions to $\dot{P}_{\rm obs}$ from an anomalous acceleration, such as from a nearby star (see next subsection), strengthening the conclusion that the latter is dominated by $\dot{P}_{\rm b, int}$.

\subsection{A triple system?}

In the timing, we measure an apparently significant second spin frequency derivative,
$\ddot{f}=-1.08\pm0.10 \times 10^{-27}\, \rm Hz\, s^{-2}$. In globular clusters, this is generally due to variations of the acceleration of the system along the line of sight; normally designated as ``jerks".
Following \cite{Freire2001,Freire2017}, we check whether this ``jerk" is mainly due to the GC potential or the gravity of its nearby stars.
The most extreme derivative of LOS acceleration from GC gravitational potential is given by:
\begin{equation}
\dot{a}_{\ell, \rm GC, max} \, = \, - \frac{3 v_{\mu,0}^2}{r_{\rm c}^2} v_{\ell, \rm max},
\end{equation}
where $v_{0}$ is the 1-D stellar velocity dispersion at the centre ($\sim 6.5 \, \rm km\,s^{-1}$)\footnote{\url{https://people.smp.uq.edu.au/HolgerBaumgardt/globular/}}, $r_{\rm c}$ is the core radius ($0.35\arcmin$ at $d = 18.5 \, \rm kpc = 1.88 \, pc$), $v_{\ell, \rm max}$ is the maximum LOS velocity of the cluster (here we assumed to be the escape velocity
25.9\,$\rm km\,s^{-1}$ of M53). These values give an estimation of $|a_{\ell, \rm GC, max}|=1.04 \times 10^{-21}\,\rm m\,s^{-3}$.
However, the observed LOS jerk, $\dot{a}_{\ell} \simeq -\ddot{f}/f \, c = 1.07(10)\times 10^{-20} \,\rm m\,s^{-3}$, is larger by one order of magnitude.

This discrepancy suggests that M53A may experience perturbations from a nearby stellar or planetary companion, despite the cluster's low density. If this is true, it would make M53A similar to PSR~J1620$-$26A, a pulsar - He WD system with $P = 11.8 \, \rm ms$, $P_{\rm b} = 191$ d and $e = 0.0253$ located in the low-density GC M4 (NGC 6121) that is thought to have a planetary companion with $\sim$1 Jupiter mass in a $\sim$ 100-yr orbit  \citep{Sigurdsson2003}. For the latter system, $\dot{a}_{\ell} \simeq -\ddot{f}/f \, c = 6.40\times 10^{-17} \,\rm m\,s^{-3}$, which is $\sim$6000 times larger than what we observe in M53A; this suggests that if the M53A system has a distant companion, it is either much less massive, or is more distant, or a combination of both. Alternatively, the observed jerk could also result from dispersion measure variations over the extended timing baseline to 35\,yr, which in the absence of multi-frequency ToAs for the early Arecibo data could masquerade as a spin frequency derivative. 

\subsection{Orbital eccentricity}

\begin{figure}
	\includegraphics[width=\columnwidth]{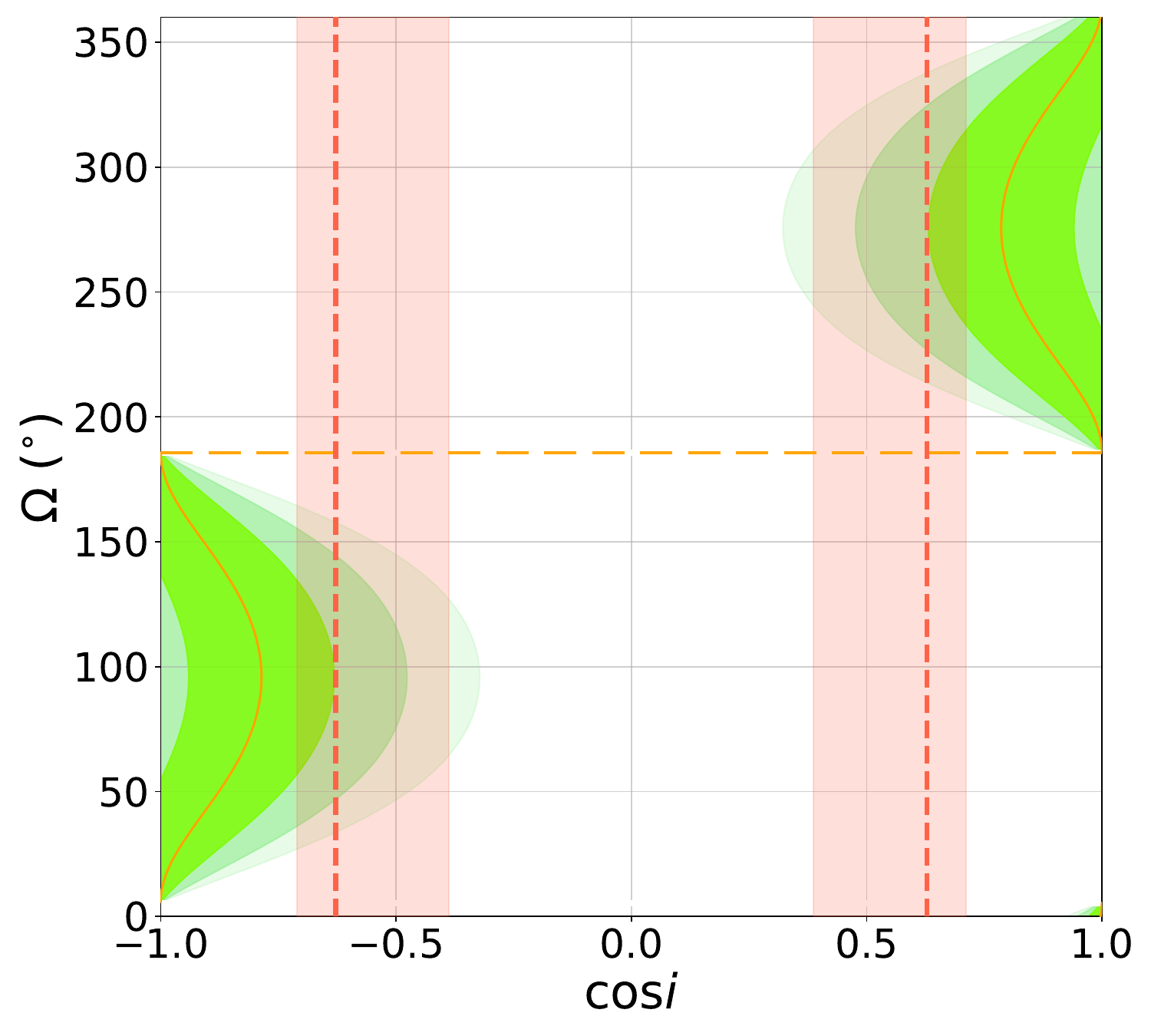}
	\caption{Orbital orientation constraints for M53A in the full $\cos i$-$\Omega$ plane. The orange solid line and the green-shaded region display the observed $\dot{x}$ and its 1, 2, and 3$\sigma$ error, respectively. The dashed orange line indicates the PA of the proper motion of M53A (185.71$^{\circ}$). The red-dashed lines with its red-shaded region indicate an orbital inclination of $51^\circ \, ^{+16}_{-6}$ (or $180^\circ - 51^\circ \, ^{+16}_{-6} = 129^\circ \, ^{+8}_{-16}$) derived from the mass function assuming the pulsar mass of 1.3\,$\Msun$ and a companion mass of $0.39^{+0.05}_{-0.07} \,\Msun$ derived from the optical photometry in Section~\ref{sec:optical counterpart}.}
        \label{xdot}
\end{figure} 

The timing solution of M53A has allowed, for the first time, a (very) precise estimate of its orbital a mildly eccentricity, $e = 0.00055732(4)$. 
Any initial eccentricity in the progenitors of binary radio pulsars dissipates efficiently due to tidal circularization during the LMXB phase, as seen in binaries in the Galactic disk, for which the eccentricity is given by \citep{Phinney1992}
\begin{equation}
\langle e^2 \rangle^{1/2} \simeq 1.5 \times 10^{-4} \frac{P_b}{100\,\mathrm{d}},
\label{epb}
\end{equation}
where $P_{\rm b}$ is the orbital period in days. M53A's eccentricity aligns with the prediction from Eq.~(\ref{epb}) ($e \sim 0.0004$), suggesting an evolutionary history similar to Galactic disk binary systems. 
This can happen in low-density GCs, where the perturbations by nearby stars have negligible effects on the orbital eccentricity, such as the binary pulsars in M13, all of which have nearly circular orbits ($<0.001$) \citep{Wang2020}. In contrast, binary MSPs in denser clusters, such as PSR J1748$-$2446ap in Terzan 5 with $e>0.9$ \citep{Padmanabh2024}, exhibit higher eccentricities due to gravitational interactions with nearby stars after the accretion ceases \citep{Rasio1995, Heggie1996}.

Following \citet{Rasio1995}, we could estimate the time scale to produce the mild eccentricity of M53A (by assuming it started with $e = 0$)
\begin{eqnarray}
\nonumber t_{>e} &&\simeq 4 \times 10^{11}\; \yr 
\left (\frac{n}{10^4\; \pc^{-3}} \right )^{-1} 
\left (\frac{v_{0}}{10\; \km\, \ps} \right ) \\
&&\times\left (\frac{P\rmsub{b}}{\mathrm{days}} \right)^{-2/3}
e^{2/5},
\end{eqnarray}
where $n$ is the number density of stars ($n \propto \rho_c$, $\rho_c$ is the density of GC), 
and $P\rmsub{b}$ is the orbital period.
Normalized with the values $\rho_c \sim 1.95\times 10^5 \; \rm L_{\odot} \, \rm pc^{-3}$ and $n \approx 4.7 \times 10^5 \; \rm pc^{-3}$ of NGC 6517 \citep{Lynch2011}, the number density $n$ is roughly estimated as $n \approx 2.6 \times 10^3 \, \rm  pc^{-3}$. These values imply a rough time of $t_{>e} \approx 1.23$\,Gyr. This aligns well with the characteristic age for M53A ($< 0.85\, \rm Gyr$), further indicating that M53A is relatively young compared to M53B to E (all larger than 2\,Gyr; \citealt{Lian2023}).
In particular, the number implies that if M53A were significantly older than indicated by its characteristic age, it should be more eccentric than observed. An example of this is M53B: it is likely much older than M53A ($\tau_c >20.45$\,Gyr), and despite its shorter orbital period (47.7\,d), it has a significantly higher eccentricity, $e = 0.013$, the largest among the known pulsars in M53 \citep{Lian2023}.

\subsection{Secular Change of the Projected Semimajor Axis}

Table~\ref{table:timingM53A} includes a statistically significant measurement of the change in the projected semi-major axis $\dot{x}=-0.022(7) \times 10^{-12}$\,lt-s\,s$^{-1}$, thus $(\dot{x}/x)^{\rm obs} = -2.6(8) \times 10^{-16} \, \rm s^{-1}$.
The observed $\dot{x}$ can arise from changes in the orbit's physical size or from the inclination altering due to the binary motion, affected by Post-Keplerian (PK) effects like the orbital decay from gravitational wave emission, Doppler modulation, spin-orbit coupling (e.g., \citealt{Lorimer2004,Stovall2019}).
However, these PK effects are negligible compared to the observed value $(\dot{x}/x)^{\rm obs}$, which can be expressed as \citep{Kopeikin1996,Stovall2019}
\begin{equation}
\left( \frac{\dot{x}}{x} \right)^{\rm obs} = \left( \frac{\dot{x}}{x} \right)^{\rm k} = \mu \, \cot i \, \sin (\Theta_{\mu}-\Omega),
\label{eq:xdot}
\end{equation}
where $\mu$ is the proper motion of the binary in mas\,$\rm yr^{-1}$, $i$ is the inclination angle, $\Theta_{\mu}$ is the position angle (PA) of the proper motion of the system ($\Theta_{\mu}=185.71^{\circ}(1)$), and $\Omega$ is the longitude of ascending node.
Using Eq.~(\ref{eq:xdot}), we display the orbital orientation constraints from $(\dot{x}/x)^{\rm obs}$ for M53A in Fig.~\ref{xdot}. The 1-$\sigma$ upper limit on $i$ derived from $\dot{x}$ is consistent with the range of inclinations derived from the optical mass estimates (Section~\ref{sec:optical counterpart}).

\begin{figure}
\centering
	\includegraphics[width=\columnwidth]{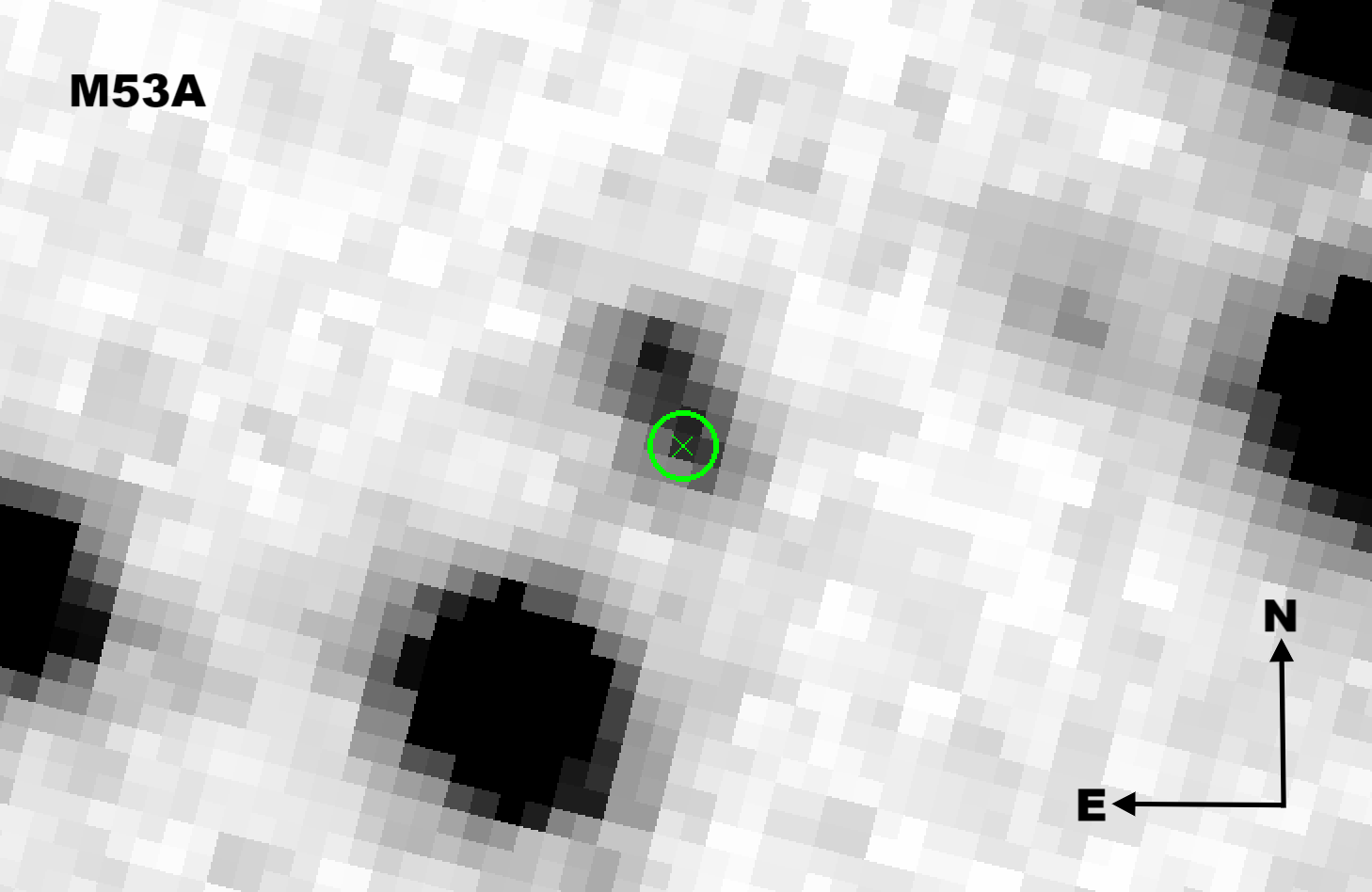}
	\caption{1$\arcsec$ $\times$ 0.7$\arcsec$ finding charts of the regions surrounding the positions of M53A in the F275W filter. In this panel, the green cross and circle indicate the pulsar position shown in Table~\ref{table:timingM53A} and a radius of 3$\sigma$ ($\sim$ 0.051\arcsec), using the optical uncertainty due to the cross-correlation between HST and Gaia DR3.}
        \label{counterpart}
\end{figure}

\begin{figure*}
\centering
	\includegraphics[width=1.5\columnwidth]{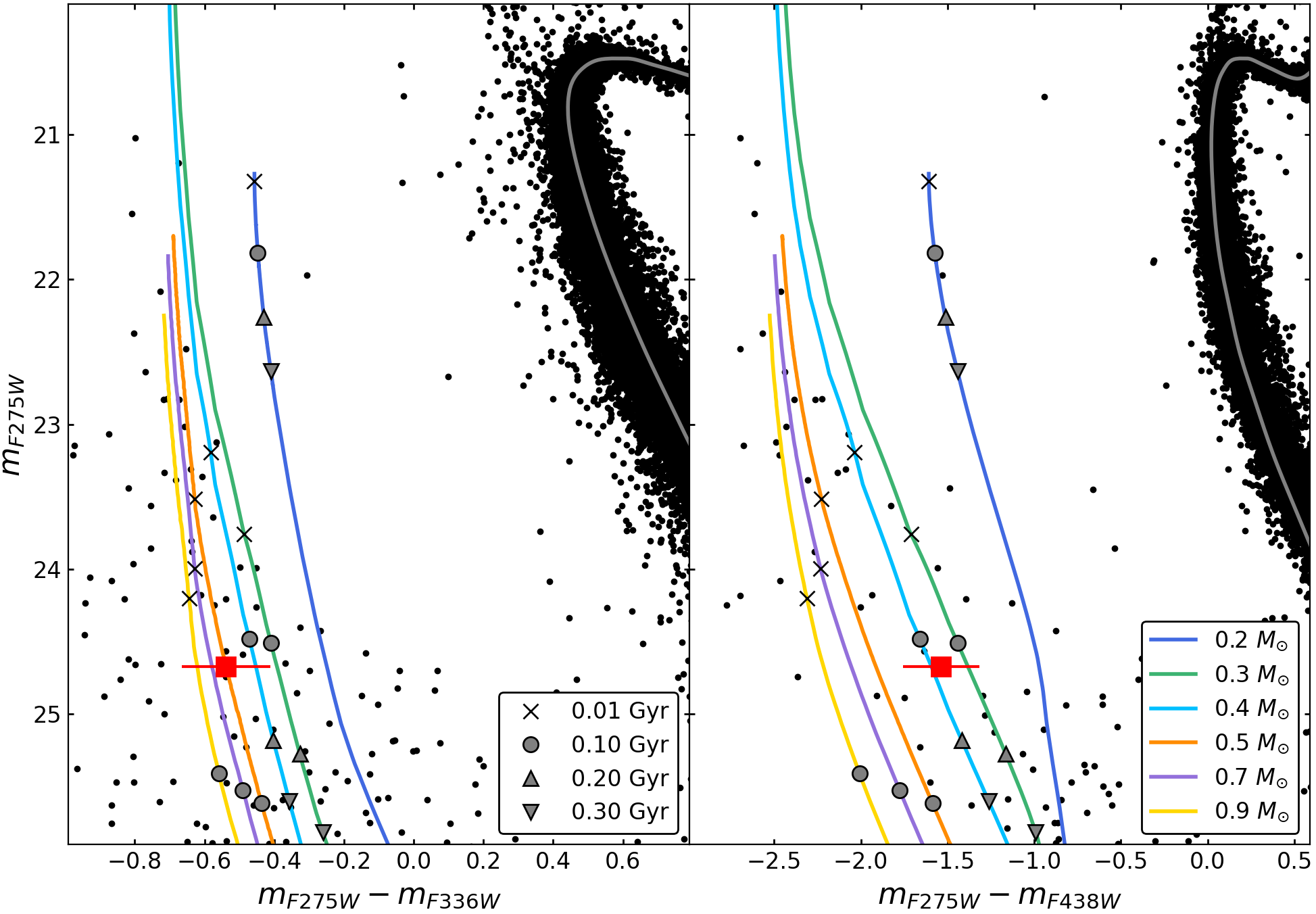}
	\caption{Color-magnitude diagram of M53 in a combination of the F275W, F336W and F438W filters. The red square is the position of the counterpart to M53A. The gray curve is a 13 Gyr isochrone calculated at the cluster metallicity, distance and extinction. The colored curves are cooling tracks for CO WD taken from than {\sc BaSTI} database \citep{Salaris2010} and He WDs from \citep{Istrate2014,Istrate2016}. The tracks are for WD with masses of 0.9, 0.7, 0.5, 0.4, 0.3, and 0.2\,$\Msun$, with decreasing masses from left to right, as reported in the legend. Different points are also highlighted with different markers along the tracks corresponding to different cooling ages.}
        \label{companion}
\end{figure*}

\section{Optical counterpart to M53A}
\label{sec:optical counterpart}

\begin{figure}
	\includegraphics[width=\columnwidth]{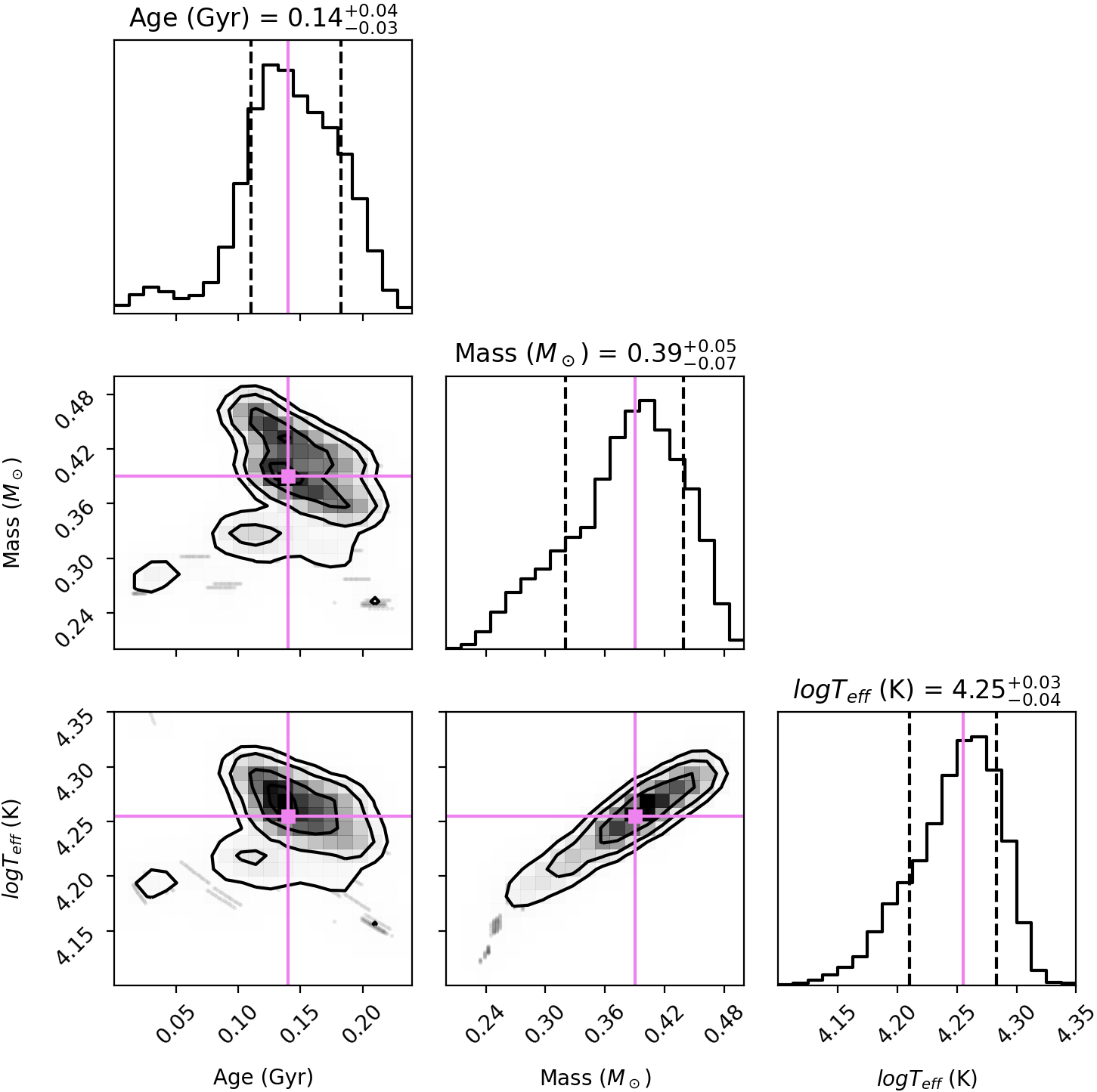}
	\caption{1D probability distributions and 2D confidence contours (at 1$\sigma$, 2$\sigma$, and 3$\sigma$ levels) for the cooling age, mass, and surface temperature of M53A's companion star. The pink solid line represents the median (50th percentile), while the black dashed lines mark the 16th and 84th percentiles, providing the best-fit estimates and associated uncertainties.}
        \label{corner}
\end{figure}

To identify the optical counterpart of M53A, we inspected all the stellar sources within a 1$\arcsec$ $\times$ 1$\arcsec$ region surrounding the pulsar (listed in Table~\ref{table:timingM53A}). The closest star is displaced by only 0.007$\arcsec$ from the timing position of M53A, with its finding chart shown in Fig.~\ref{counterpart}. In the color-magnitude diagram (CMD), the star is located along the WD sequence (see Figure~\ref{companion}), as expected from the binary orbital properties of the pulsar and from the standard MSP formation scenario. The excellent positional coincidence and the CMD location provide strong evidence that the detected source is the companion to M53A. There is a second star very close to the pulsar position, but at a larger distance of 0.07\,arcsec from the pulsar position and along the main sequence in the CMD, suggesting that it is unlikely to be the companion to M53A. The companion magnitudes are: $m_{F275W}=24.67\pm0.04$, $m_{F336W}=25.21\pm0.12$, and $m_{F438W}=24.2\pm0.2$. M53 has also been observed with HST's Advanced Camera for Surveys in optical filters. However, due to the lower angular resolution of the camera with respect to UVIS/WFC3, the companion is not detectable because the flux is dominated by the nearby main sequence star.

Using the same method outlined in \citet{Cadelano2019,Cadelano2020}, we exploited the multi-band photometry of the counterpart to infer the physical properties of the companion.
We compared the measured magnitudes with those predicted by carbon-oxygen (CO) WD \citep{Salaris2010} and helium (He) WD cooling sequences \citep{Istrate2014,Istrate2016}. The corresponding curves are shown in Fig.~\ref{companion}, together with a 13\,Gyr isochrone (gray solid curve) extracted from the BaSTI database \citep{hidalgo2018,pietrinferni21} at the cluster metallicity $[Fe/H]=-2.1$ (gray solid curve). For both the isochrone and the WD cooling models, we assumed a cluster distance modulus of $(m-M)_0=16.32$ and a color excess of $E(B-V)=0.02$, in very good agreement with the values quoted by \citet{Harris1996,Harris2010}.

To constrain the companion mass, effective temperature and cooling age we calculated for each point of each cooling track the likelihood based on the differences between the observed magnitudes and the model-predicted magnitudes. For each point, we computed the difference
\[
\Delta m = m_{\text{obs}} - m_{\text{model}},
\]
where \(m_{\text{obs}} = (m_{F275W}, m_{F336W}, m_{F438W})\) are the observed magnitudes, and \(m_{\text{model}}\) is the same but for model-predicted magnitudes. The likelihood for each model was computed as
\[
\mathcal{L} \propto \exp\left(-\frac{1}{2} \chi^2\right), \quad \text{where} \quad \chi^2 = \sum_{i} \frac{\left(\Delta m_i\right)^2}{\sigma_{\text{obs}, i}^2}.
\] 
Here, \(\sigma_{\text{obs}, i}\) are the observational uncertainties associated with each magnitude. The likelihood distribution was normalized, and the best-fit parameters were determined from the marginalized distributions by using the 0.16th, 0.50th, 0.84th percentiles. The best-fit values (see Fig.~\ref{corner}) are: cooling age of $0.14^{+0.04}_{-0.03}\, \rm Gyr$, mass $M_{\rm WD}=0.39^{+0.05}_{-0.07} \, \Msun$, and effective temperature of $T = (18 \pm 1) \times 10^3\, \rm K$. The inferred mass is compatible with the companion being a He WD rather than a CO one. The companion mass agrees well with the prediction for WDs formed from Population II stars (the metallicity $\rm Z=0.001$; the hydrogen fraction $\rm X=0.75$), with an initial mass of 1.0\,$\Msun$ leading to a final mass of 0.41\,$\Msun$ in a 256-day binary system \citep{Tauris1999}.
This is also consistent with the lower limit of the companion mass $M_{\rm c,min} = 0.31\,\Msun$ derived from the mass function in Table~\ref{table:timingM53A}.

Assuming a pulsar mass of 1.3\,$\Msun$ and a companion mass of $0.39_{-0.07}^{+0.05}\,\Msun$, we obtain from the mass function an orbital inclination of $i = 51^{\circ} \, _{-6}^{+16}$ or 
$i = 180^{\circ} - 51^{\circ} \, _{-6}^{+16} = 129^{\circ} \, _{-16}^{+8}$. These constraints are consistent with the limits on orbital inclination derived from our measurement of $\dot{x}$ (see Fig.~\ref{xdot}), especially for the larger companion masses within this range, like the \cite{Tauris1999} prediction of $0.41\, \Msun$ for a He WD, and/or lower pulsar masses, as these result in lower orbital inclinations.

\begin{figure*}
\centering
\label{ppdot}
	\includegraphics[width=0.6\textwidth]{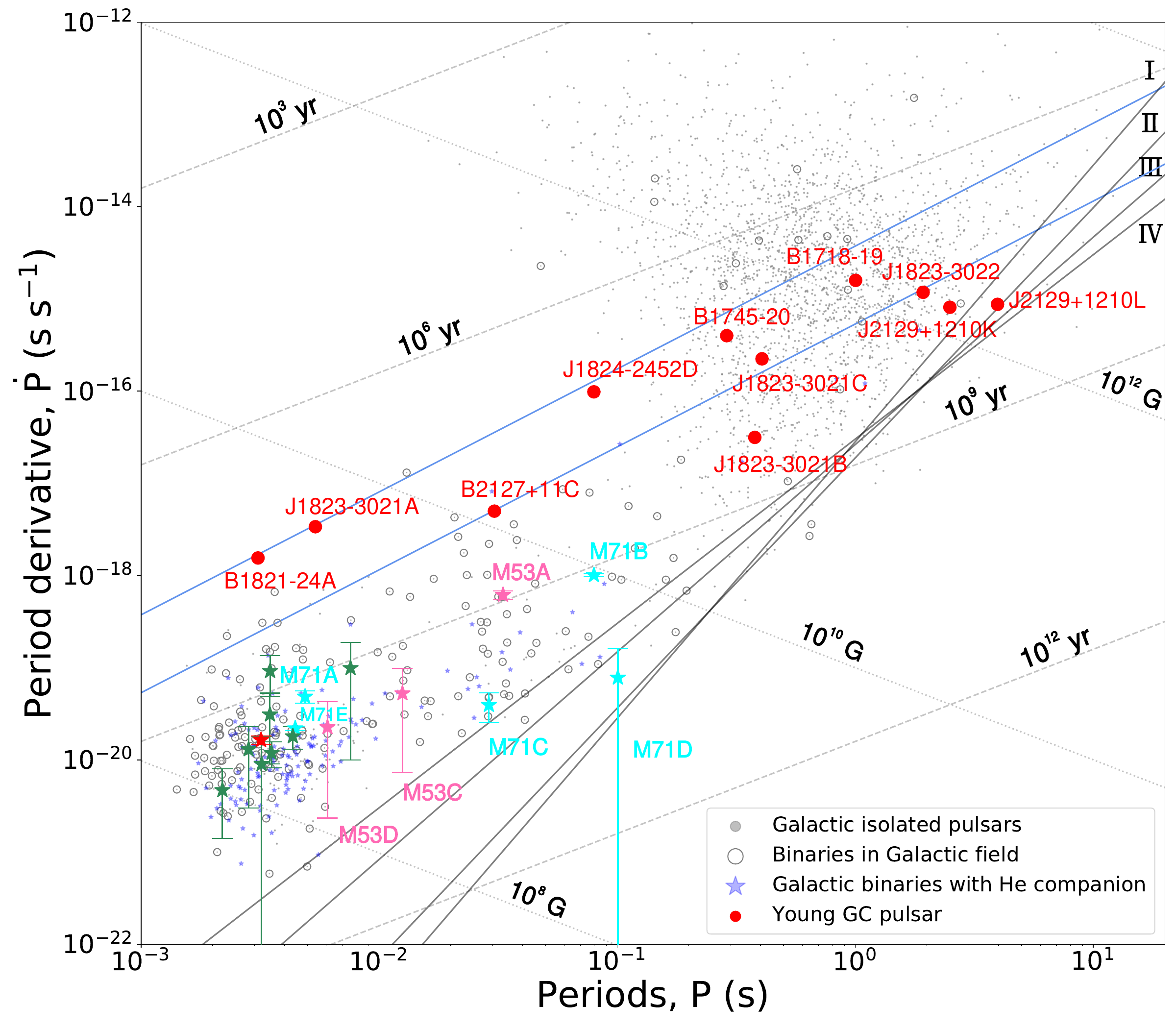}
	\caption{Period–period derivative plot for the pulsars in the ATNF Pulsar Catalog. The blue solid lines denote the spin-up line according to \citet{Verbunt2014} (see detailed discussion in \citealt{Tauris2012}) and four models of the death line are marked by solid black lines from I to IV following \citet{Cruces2021}. The derived period derivatives for the MSPs in 47 Tuc (shown as green stars), M53 (shown as the pink stars), M71 (shown as the cyan stars; Y. Lian et al. in prep.), and NGC 6749A (shown as the red star; P. Freire et al. in prep.) place them in the same region of the diagram where similar system in the Galactic disk occur; these clusters do not seem to have young pulsars. The young ($\tau_c < 10^8$ yr) pulsars in denser clusters are indicated by the red dots and labels \citep{Abbate2022,Zhou2024,Wu2024}.}
    \label{fig:ppdot}
\end{figure*}

\section{Discussion and conclusions}
\label{sec:conclusions} 

\subsection{A consistent picture for M53A}

Using the WD mass from Section \ref{sec:optical counterpart}, we estimated the proto-WD phase duration (time from Roche-lobe detachment to reaching the WD cooling track; see \citet{Istrate2014}), finding $\sim$0.2\,Gyr. Adding this to the cooling age, the total age is $\sim$0.34\,Gyr, which is consistent with M53A’s characteristic age ($<0.85$\,Gyr) and the orbital eccentricity timescale $t_{>e} \approx 1.23$\,Gyr mentioned above. In particular, this age represents only 2.8 \% of the age of M53. These results confirm that the companion of M53A is a relatively young He WD, formed within the last few percent of the history of M53.

The mildly recycled nature of M53A ($P \sim$33\,ms, $B = 4.5$ - $5.0 \times 10^9 \, \rm G$) and its mild eccentricity make it analogous to wide Galactic pulsar–He WD systems. All these systems form in case B Roche lobe overflow \citep{Tauris2023}, which happens after hydrogen burning ceases in the WD progenitor, and the star enter the giant branch. Among such systems, the widest for which we find spin periods below 10\,ms is PSR~J1640+2224, with $P_{\rm b}\, = \,175$~d and $P\, = \, 3.3$\,ms; beyond this, all pulsars in these systems have, like M53A, significantly slower spin periods and larger magnetic fields \citep[see discussion in section 14.2 of][]{Tauris2023}. This suggests that in these systems mass transfer began when the donor star was near the tip of the giant branch and expanding rapidly. Their envelopes likely transferred quickly at a super-Eddington rate, resulting in inefficient spin-up and less ablation of the magnetic field of the NS \citep{Rappaport1995} (see Fig.~\ref{fig:ppdot}), compared with normal MSPs (e.g. M53C, D, and E).

This also makes the system similar to other wide, low-eccentricity systems seen in low-density GCs, like the aforementioned 191-d binary pulsar PSR~B1626$-$26A in M4 \citep[where, as in the case of M53A, the WD companion has been optically identified as $0.48 \pm 0.14$ Gyr, low-mass WD,][]{Sigurdsson2003} and M71B with $P = 79.9 \, \rm ms$, $P_{\rm b}\sim 466 \,\rm d$ and M71C with $P = 28.9 \, \rm ms$, $P_{\rm b}\sim 378 \,\rm d$ (see details in Lian et al., in prep.).
These findings confirm the theoretical expectation that wide, nearly circular systems have a much higher chance of survival in low-density globular clusters, which are much unlike the dynamically active environments of denser clusters (e.g., the many eccentric pulsars in NGC~1851 or Terzan~5; \citealt{Ridolfi2022,Padmanabh2024}).

The special characteristics of this system (being located in a GC, having a WD identification) allow us to learn something about this type of system. The optically
derived age estimate of 0.35 Gyr allows an estimate of the spin period after accretion ceased. The spin period of a pulsar as a function of time is given by \citep{Lazarus2014}:
\begin{equation}
\frac{p(t)}{p} = \left[1 +  \frac{ (n-1) \dot{p}}{p} t  \right]^{\frac{1}{n -1}},
\end{equation}
where $n$ is the braking index, which in this equation must be larger than 1. For $n$ = 2,3 and 4 and $t = -0.35$ Gyr, the values of $p(t)$ are, respectively, 25.7, 24.6 and 22.9 ms. Finally, the age of the GC is such that a $\sim\,1\, \rm M_{\odot}$, or even slightly lower, main sequence progenitor of this WD should have just left the main sequence and become a WD.

\subsection{Slow Pulsars in Low and High-Density GCs}

In the discussion on slow pulsars in GCs, \cite{Verbunt2014} stated that slow pulsars occur exclusively in high-density GCs. In low-density GCs, LMXBs evolve undisturbed towards their final evolutionary state, an MSP with a low-mass companion in a low-eccentricity orbit. In denser GCs, LMXBs could be disrupted, leaving behind slow, partially recycled pulsars \citep{Verbunt2014}. Note, however, that there are other proposed formation paths for these pulsars: \cite{Kremer2024} argue, from N-body simulations, that WD mergers could also effectively produce slow pulsars in GCs, \cite{Tauris2013} suggest the possibility of accretion-induced collapse of WDs and \cite{Lyne1996} suggest direct collisions with a main sequence stars, see general discussion in \cite{Boyles2011}.

Some of the recent discoveries in the low-density GC M71 (Lian et al., in prep.) are, like M53A, relatively slow-spinning pulsars. 
As discussed, this is likely because they are what their orbital characteristics suggest: wide slow pulsar - He WDs similar to those of the Galactic disk; these systems survived in these clusters owing to their much lower stellar densities. Thus, they are presumably different from the slow pulsars in the very dense GCs.

To verify this, we have placed these pulsars in the $P$-$\dot{P}$ diagram, see Fig.~\ref{fig:ppdot}.
The systems indicated with the large red dots are in high-density GCs, these are listed by \cite{Abbate2022}, to these we add two slow pulsars in M15 \citep{Zhou2024,Wu2024}. We can place them in this diagram because their positive values of $\dot{P}_{\rm int}$ are much larger than the effect of the relatively large accelerations caused by the gravitational fields of their host GCs, which otherwise generally make estimates of  $\dot{P}_{\rm int}$ impossible. In addition to these, we also place the pulsars in M53 and M71 in this diagram (in cyan and blue); we can do this because, in these GCs, the predicted cluster accelerations are much smaller, this means that even moderate values of $\dot{P}_{\rm int}$, like that of M53A, can be estimated with astrophysically useful precision, as discussed in Section~\ref{sec:period_derivatives}. Finally, we add some of the estimates of $\dot{P}_{\rm int}$ for the pulsars in 47~Tuc (in green) and NGC 6749 (in red), where independent measurements of acceleration effects have been made using orbital period derivatives \cite[][Freire et al. in prep.]{Freire2017}, as we have done for M53A in Section~\ref{sec:period_derivatives}.

The conclusions are clear. Firstly, the observed $\dot{P}_{\rm int}$ are all below the adjusted spin-up line discussed by \cite{Verbunt2014} (for the rationale for this adjustment, see \citealt{Tauris2012}) which is consistent with partial accretion and disruption; this is also true for the ``young" pulsars found since, as discussed by \cite{Abbate2022} and confirmed in the new discoveries of \cite{Zhou2024,Wu2024}\footnote{However, not all slow pulsars in dense clusters are that young: if they have small values of $\dot{P}_{\rm int}$, we won't be able to estimate them because of the large accelerations predicted by these clusters.}. If they were genuinely new NSs, as suggested e.g., by \cite{Kremer2023}, we should be able to find pulsars in GCs above this line. However, the value of the spin-up lines are subject to uncertainties \citep{Tauris2012}. Some slow pulsars appear above the non-adjusted spin-up line (see the lower blue solid line in Fig.~\ref{fig:ppdot}), suggesting that something unusual (as discussed above) is happening in the denser GCs; furthermore, the absence of such detections does not rule out WD mergers or accretion-induced collapse, as spin-down may occur rapidly. Secondly, their proximity to the spin-up line (many have characteristic ages smaller than $100\,\rm Myr$, which is less than 1\% of typical GC ages) indicates that in some of the densest GCs the process forming these apparently young pulsars is ongoing. Thirdly, it is also clear that these ``young" pulsars are absent in the low-density GCs, like M53 and M71 (and the MSPs in 47~Tuc and NGC~6749); the latter - both fast and slow - are very similar, in their spin-down rates, to similar systems in the disk of the Galaxy.
Future discoveries of additional slow pulsars in GCs will further constrain their formation mechanisms.

\subsection{Conclusions}

M53 is the most distant cluster with known pulsars, and M53A is one of the first pulsars ever found in a GC. It is a 33\,ms pulsar in a $\sim 256\,\rm d$, low-eccentricity orbit with a low-mass companion. However, until now (35 years after its discovery), no timing solution had ever been published. Thanks to the high sensitivity of Arecibo and FAST, we have obtained a timing solution for M53A with a 35-year baseline from 1989.03 to 2024.05, and an optical identification in HST data, which provides a consistent understanding of this system.

\begin{itemize} 

\item A precise position and proper motion. The pulsar lies 0.52 arcminutes (1.5 core radii) from the center of the cluster. The proper motion of this pulsar measured in this work, $\mu_{\alpha}\, =\, -0.36 \, \rm mas\,yr^{-1}$ and $\mu_{\delta}\, =\, -0.62 \, \rm mas\,yr^{-1}$, has a 3.7-$\sigma$ difference from the cluster proper motion measured from the Gaia DR3; the real difference is likely much smaller; and we might have some unknown systematics in our timing.

\item The measurement of the pulsar's spin-down, M53A has a relatively large $\dot{P}_{\rm obs}$ that cannot be explained by acceleration in the cluster, this is clearly dominated by the intrinsic spin-down of the pulsar. The limits on $\dot{P}_{\rm int}$ result in a characteristic age between 0.70 and $0.85\,\rm Gyr$ and a low $B$ field (between 4.5 and $5.0 \times 10^{9} \, \rm G$). This and the orbital eccentricity of $\sim$0.00056 (which is now precisely measured) make this system similar to wide pulsar - He WD systems in the Galaxy. If the system were much older, the orbital eccentricity would likely be significantly larger, as observed for M53B.

\item The precise position derived from the timing allowed the discovery of the optical counterpart for the companion of M53A in archival HST images. The photometry of this companion indicates that, as expected from the spin and orbital parameters, it is a fairly massive ($M_{\rm WD}=0.39^{+0.05}_{-0.07} \, \Msun$) He WD  with an effective temperature of $18 \pm 1 \times 10^3 \, \rm K$ and a cooling age of $0.14^{+0.04}_{-0.03}\, \rm Gyr$.
The mass is consistent with the minimum companion mass inferred from the mass function of the system. The cooling age, plus the time it has likely spent as a proto-WD (adding to 0.35 Gyr) is physically consistent with the characteristic age derived from timing.

\item We find that there are significant numbers of slow-spinning pulsars in the lowest-density GCs. Their spin-down rates are within the same range as similar systems observed in the Galactic disk. In high-density GCs, many pulsars, especially the slower ones, have large spin-down rates indicating either ongoing formation and disruption of LMXBs, or ongoing formation of new pulsars via other mechanisms; those systems are thus far absent in the low-density GCs.

\end{itemize} 
\vspace{0.5cm}

This work is supported by National Natural Science Foundation of China under Grants Nos. 12021003, 12433001, 12041301, 12173052, 12173053, and 11920101003; Beijing Natural Science Foundation No. 1242021; the Fundamental Research Funds for the Central Universities. P. C. C. F. gratefully acknowledges continued support from the Max-Planck-Gesellschaft, and the interesting discussions with Thomas Tauris.
Z. Pan and L. Qian are supported by the CAS ``Light of West China" Program and the Youth Innovation Promotion Association of the Chinese Academy of Sciences (ID 2023064, 2018075, and Y2022027), National Key R\&D Program of China, No. 2022YFC2205202, National SKA Program of China No. 2020SKA0120100. Y. Lian is supported by the China Scholarship Council (Grant No. 202306040148). 
This work made use of the data from FAST \footnote{\url{ https://cstr.cn/31116.02.FAST}}. FAST is a Chinese national mega-science facility, operated by National Astronomical Observatories, Chinese Academy of Sciences. We thank Alex Wolszczan, Stuart B. Anderson, Bryan Jacoby and Shrinivas Kulkarni for the earlier Arecibo measurements of the times of arrival for M53A. At the time of these observations, the Arecibo Observatory was part of the National Astronomy and Ionosphere Centre, and was operated by Cornell University under contract with the NSF.


\end{document}